\documentclass[a4paper,11pt]{article}
\usepackage{pos}

\title{Detection of low-energy X-rays with 1/2 and 1 inch LaBr3:Ce crystals read by SIPM arrays}
\ShortTitle{Detection of low-energy X-rays by LaBr$_3$:Ce crystals read by SiPM arrays}

\author*[a]{M. Bonesini}
\author[a]{R. Bertoni}
\author[a]{R. Benocci}
\author[a]{M. Clemenza}
\author[a]{R. Mazza}
\author[b]{A. deBari}
\author[b]{A. Menegolli}
\author[b]{M. Prata}
\author[b]{M. Rossella}

\affiliation[a]{Sezione INFN Milano Bicocca,Dipartimento di Fisica G. Occhialini
, Dipartimento di Scienze dell' Ambiente e della Terra, 
Universit\'a Milano Bicocca,  Piazza Scienza 3, Milano, Italy}

\affiliation[b]{Sezione INFN Pavia, Dipartimento di Fisica, Universit\'a di Pavia,  via A. Bassi 6, Pavia, Italy}

\emailAdd{Maurizio.Bonesini@mib.infn.it}

\abstract{LaBr$_3$:Ce crystals, with photomultiplier or single SiPM 
readout (up to 3x3 mm$^2$)
have been introduced for radiation imaging in medical physics.
An R$\&$D was pursued with 1/2" and 1" LaBr3:Ce crystals,
from different producers,
to realize  compact large area detectors (up to some cm $^2$ area) with SiPM
array readout,  aiming at high light yields, good energy resolution,
good detector linearity  and fast time response for low-energy X-rays.
The study was triggered by  the FAMU experiment at the RIKEN-RAL muon
facility,  aiming at a precise  measurement of the proton Zemach radius to
solve the  so-called "proton radius  puzzle". The goal is the detection of
characteristic  X-rays around 130 KeV. Other applications may be foreseen
for homeland security and $\gamma$-ray  astronomy. Results were 
obtained with a direct readout  based on a CAEN V1730 FADC, with no need for
an amplification stage.
At the Cs$^{137}$ peak, 
energy resolutions up to $\sim 3 \%$ were obtained, using  
a readout with Hamamatsu SiPM arrays.
These results  compare well with best available results  obtained with a
PMT readout.}

\FullConference{%
  *** The European Physical Society Conference on High Energy Physics (EPS-HEP2021), ***\\
  *** 26-30 July 2021 ***\\
  *** Online conference, jointly organized by Universität Hamburg and the research center DESY ***
}

\begin{document}
\maketitle

\section{Introduction}
\begin{figure}[htbp] % figures (and tables) should go top or bottom of
                    % the page where they are first cited or in
                    % subsequent pages
\centering
\includegraphics[width=.53\textwidth]{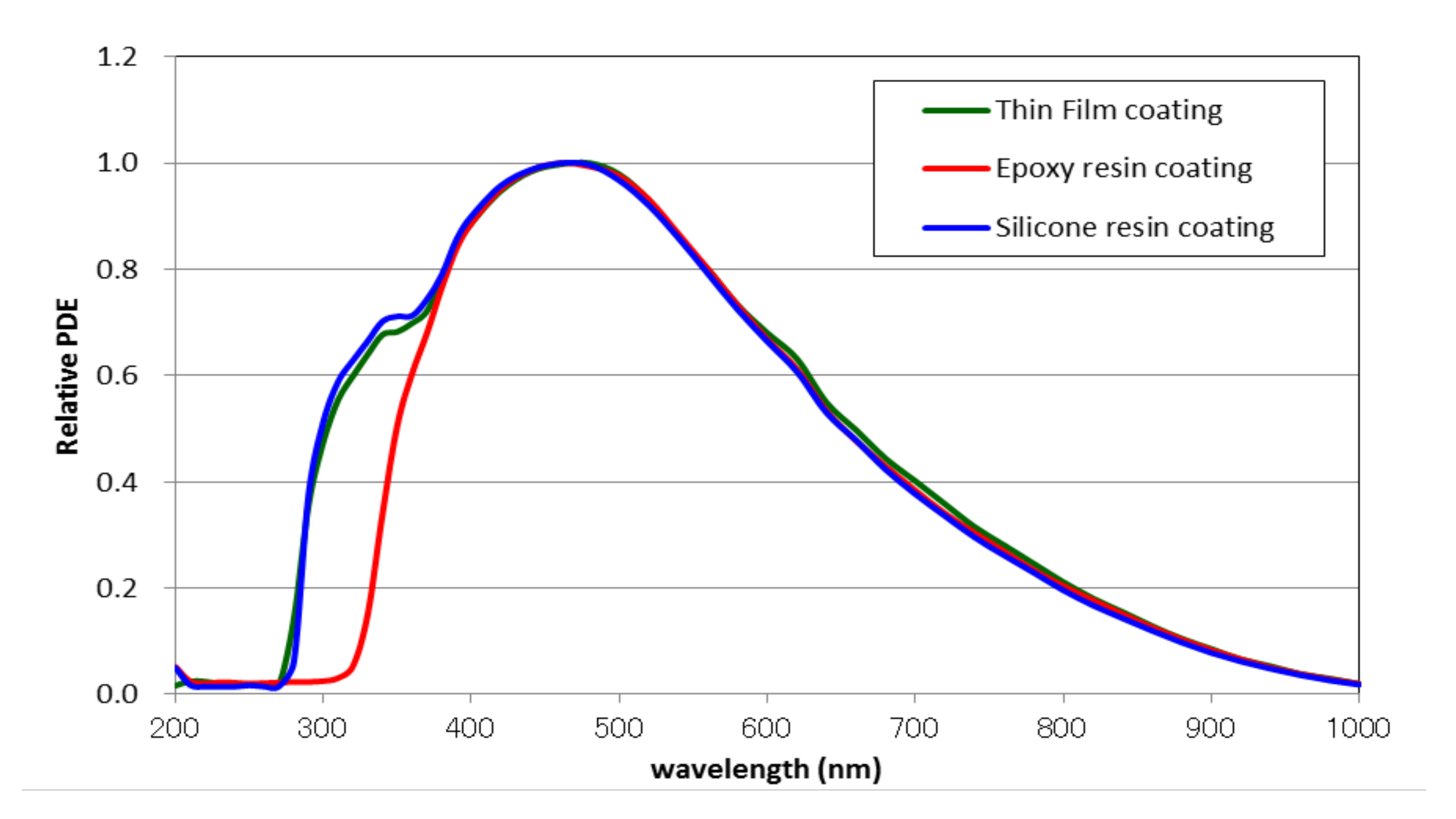}
\includegraphics[width=.36\textwidth]{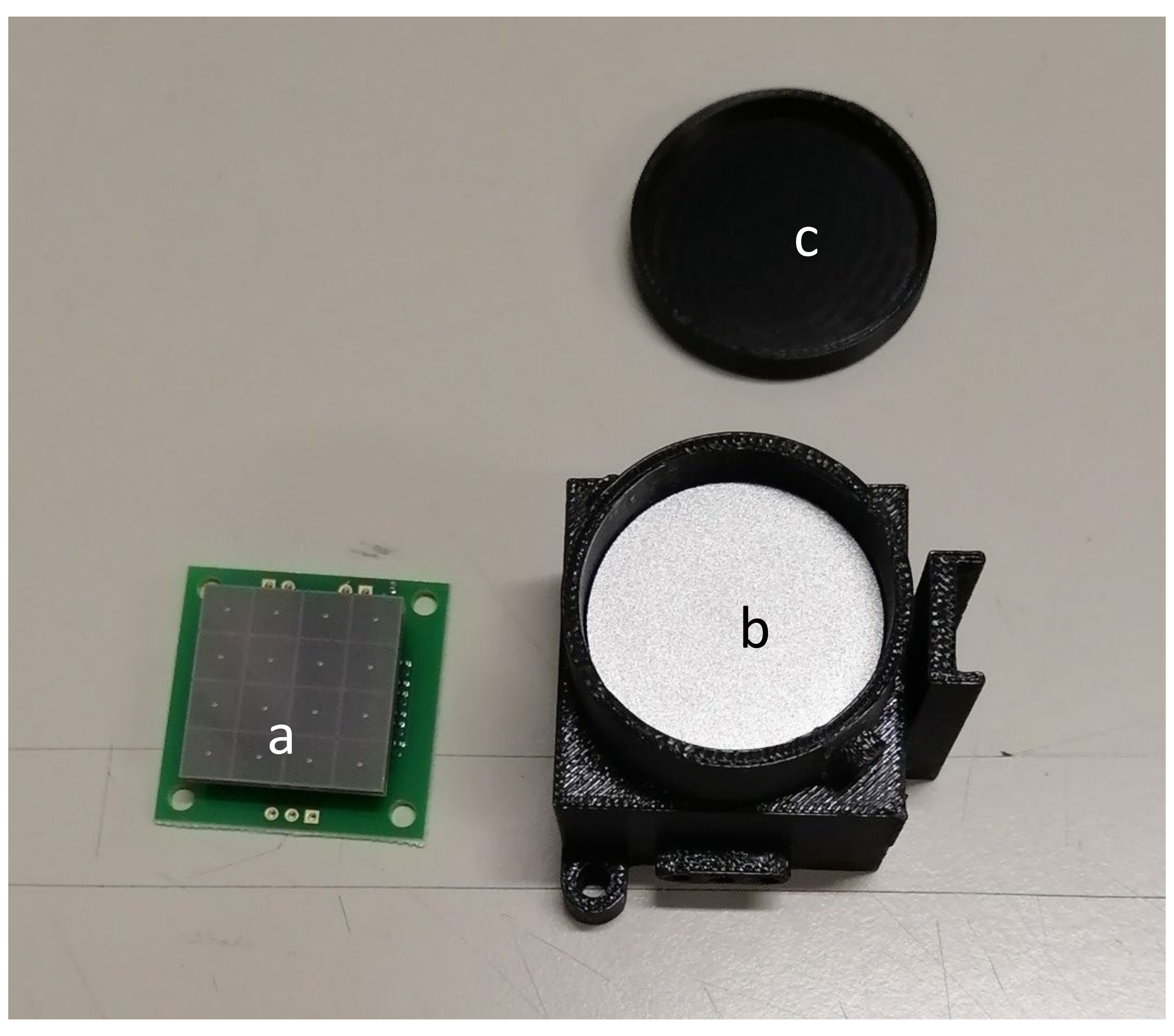}
\caption{Left panel: 
response of Hamamatsu S13361 SiPM arrays to incident photons of 
different wavelengths and with different windows  [courtesy of Hamamatsu 
Italia]. The  response in the UV region is enhanced  with a Silicone resin
coating.
Right panel:
1" detector mounting. a) SiPM array mounted on the PCB; b) crystal under test 
inside the holder; c) cover to ensure light-tightness. Mounting for 1/2" detectors is similar.}
\label{fig:det}
\end{figure}

The need of large area, compact crystal detectors, with good linearity and energy resolution, read by
SiPM is arising in many fields from fundamental physics experiments, such as FAMU \cite{mb1,cp1}, to
PET \cite{pet}, homeland security  and $\gamma-$ ray astronomy.  
Even if they are hygroscopic, the good light yeld (75000 $\gamma$/MeV) and
short decay time ($\sim 30$ ns) of LaBr$_3$:Ce crystals 
makes them the best choice, as respect
to other crystals as PrLuAg or Ce:GAGG \cite{mb4}. 
Our $R\&D$ is based on LaBr$_3$:Ce crystals of increasing size (1/2
inches to 1 inches) read by SiPM arrays, whose gain drift with temperature 
is corrected online by a custom NIM module.
The signals from the different cells of the readout array (4 for SENSL 
arrays, 16 for the other arrays) are
summed up and read in parallel, to be later digitized by a fast CAEN V1730 FADC. As the signal is
sizeable: $\sim 100$ mV at the Cs$^{137}$ peak, there is no need for an amplification stage. 
The crystal and the SiPM array are housed in a 3D ABS printed holder, as 
shown in the right panel of figure 
\ref{fig:det}. The optical contact between the crystal and the SiPM array 
is realized through 
Bicron BC631 silicone optical grease. 
To correct for the gain drift of SiPM with temperature, an Analog Devices
TMP37  sensor is mounted on
the back of each SiPM array. The temperature information is then used 
by a custom made module, based on CAEN A7585D units,  to
make an online correction (see reference \cite{mb1,mb3} for more details). The effect on the
detector response (P.H. in a.u.) between
10 $^{\circ}$C and 40 $^{\circ}$ C is reduced from 60 (40) $\%$ to less than 
6 (10) $\%$ for 1/2" (1") detectors (see figure \ref{fig:crys} for an example).
While 1/2" crystals are cubic in shape: $ 14 \times 14 \times 14$ mm$^3$, 1" crystals are round and
0.5" thick~\footnote{ we estimated a 88 $\%$ efficiency for 100 (200) keV 
X-rays 
with a 0.33 (1.54) cm thickness crystal, from tabulated X-ray mass attenuation coefficients \cite{nist}.
This estimation has been confirmed by a refined MonteCarlo simulation based 
on the MNCP code \cite{mncp}.
Thus for detection of signal X-rays with energy 
$\sim 130$ keV in FAMU, a 0.5 inch thickness 
is adequate.}
The used SiPM arrays are shown, with their main characteristics, in table
\ref{tab:sipm}.
Operating voltages $V_{op}$ are set to $V_{brk}$ + overvoltage, where the overvoltage is
chosen according to manufacturer's specifications.
\begin{table}[htb]
\caption{Main characteristics of the SiPM 1/2 " and 1" arrays used in the laboratory
tests. } 
\label{tab:sipm}
\smallskip
\centering
\begin{tabular}{|l|c|c|c|c|c|c|c|}
\hline
	& size  & cell dim & V$_{op}$ & $\Delta$V$_{brk}$/T  &  
$\lambda_{peak}$  & PDE max& spectral range \\ 
 & (inches) & (mm$^2$) & (V) & mV/C & (nm) & ($\%$) & (nm) \\ \hline
Hamamatsu           &   &       &      &      &     &       &        \\
S14161-6050-AS      & 1 & 6x6   & 41.1 & 34   & 450 & 50    & 270-900 \\ \hline
SENSL               &   &       &      &      &     &         &         \\
Array-J-60035-4P    & 1/2 & 6x6 & 29   & 21.5 & 420 & 50  & 200-900 \\ \hline
Advansid            &   &       &      &      &     &         &      \\ 
NUV3S-4x4-TD        & 1/2 & 3x3 & 29.5 & 26   & 420 &  43 & 350-900  \\ \hline
Hamamatsu           &   &       &      &      &     &         &      \\ 
S14161-3050-AS      & 1/2 & 3x3 & 41.1 & 34   & 450 &  50  & 270-900  \\ \hline
Hamamatsu           &   &       &      &      &     &         &      \\ 
S13161-3050-AS      & 1/2 & 3x3 & 53.8 & 60   & 450 &  35  & 320-900  \\ \hline

\hline
\end{tabular}
\end{table}

\section{Results from laboratory tests}
\begin{figure}[htbp] % figures (and tables) should go top or bottom of
                    % the page where they are first cited or in
                    % subsequent pages
\centering
\includegraphics[width=0.47\textwidth]{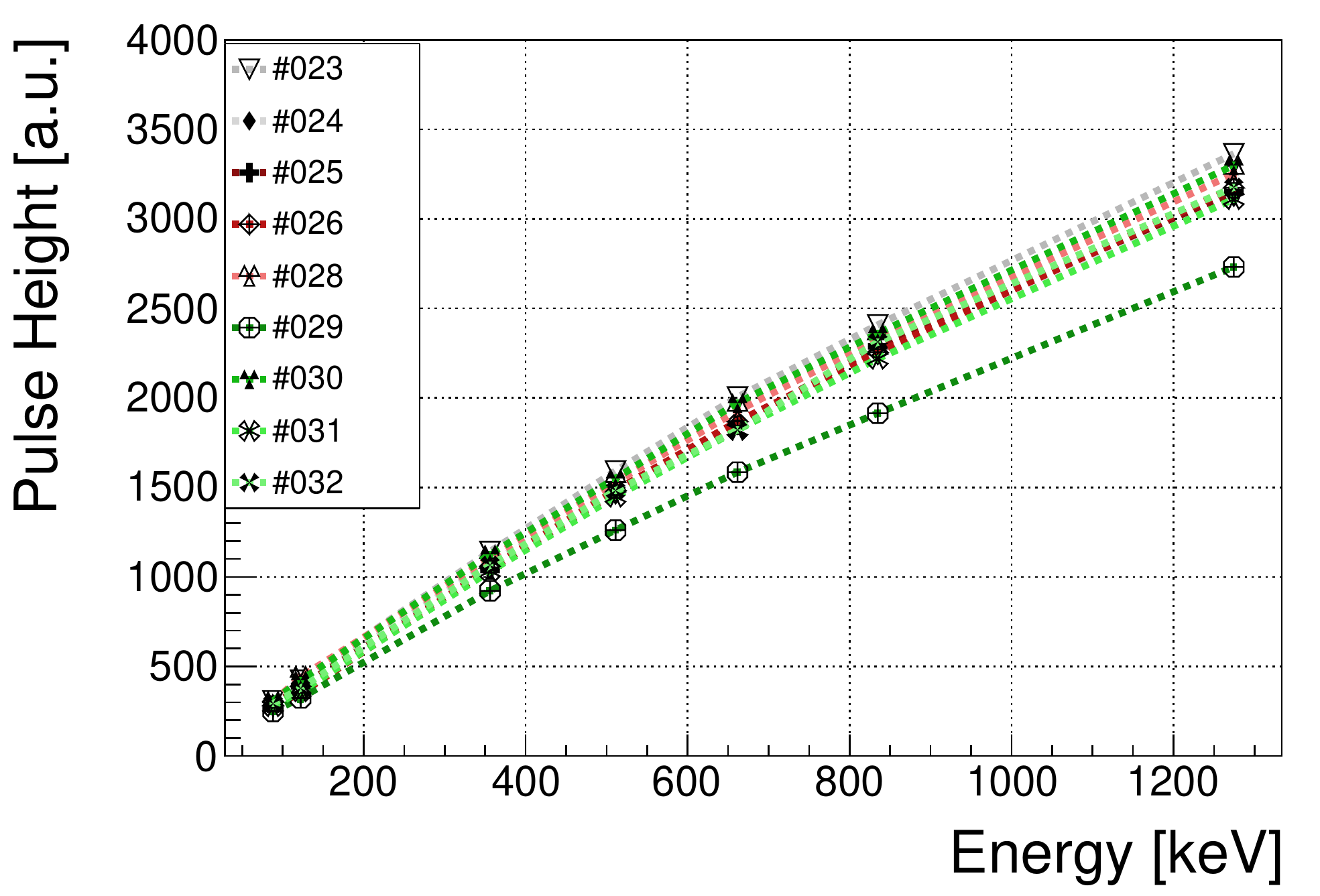}
\includegraphics[width=0.47\textwidth]{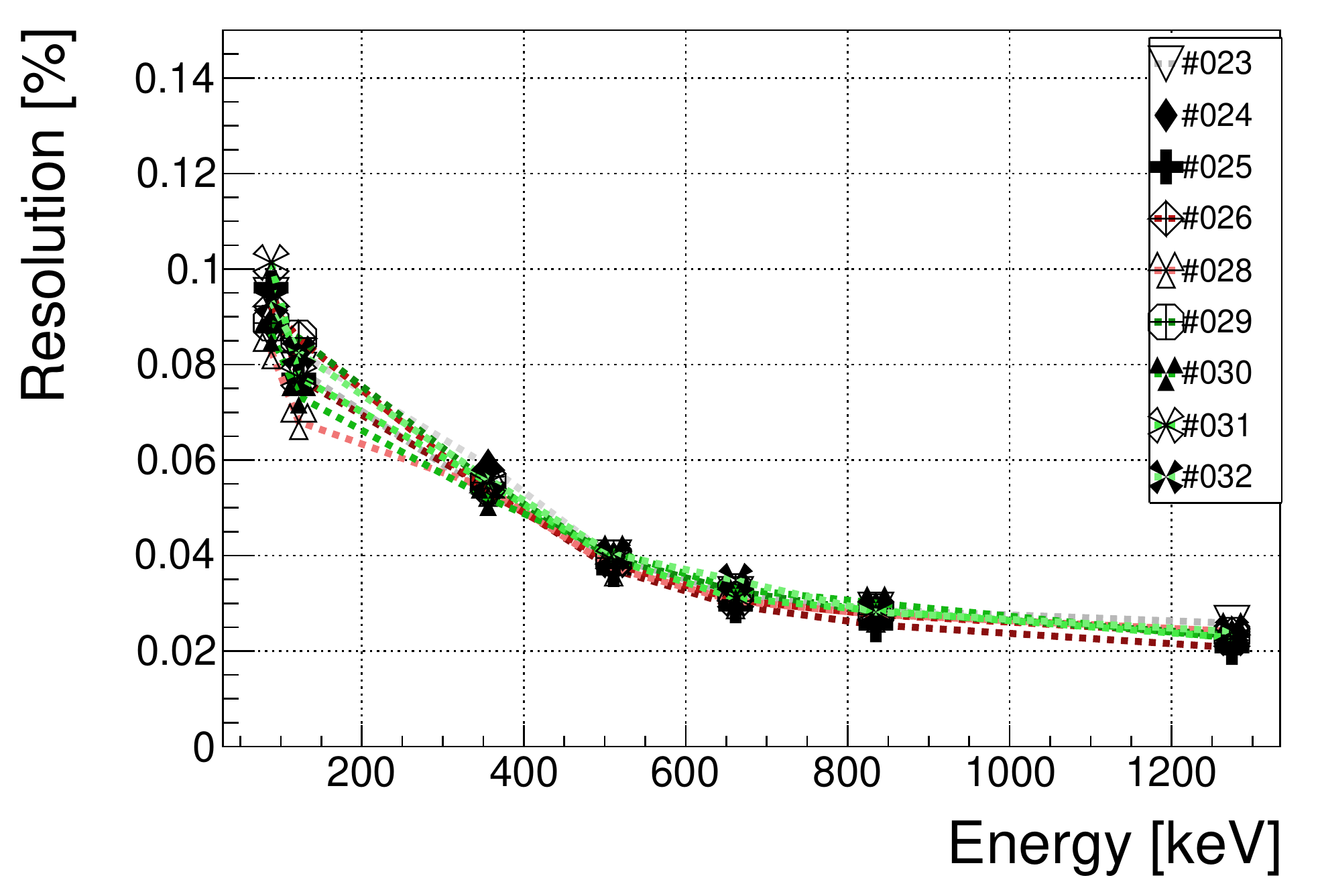}
\caption{Linearity (left panel) and energy resolution (right panel) dependence 
	for a sample of 1/2'' LaBr$_3$:Ce crystals read by Hamamatsu S13161-3050-AS 
	SiPM arrays.} 
\label{fig:crys1}
\end{figure}

\begin{figure}[htbp] % figures (and tables) should go top or bottom of
                    % the page where they are first cited or in
                    % subsequent pages
\centering
\includegraphics[width=0.47\textwidth]{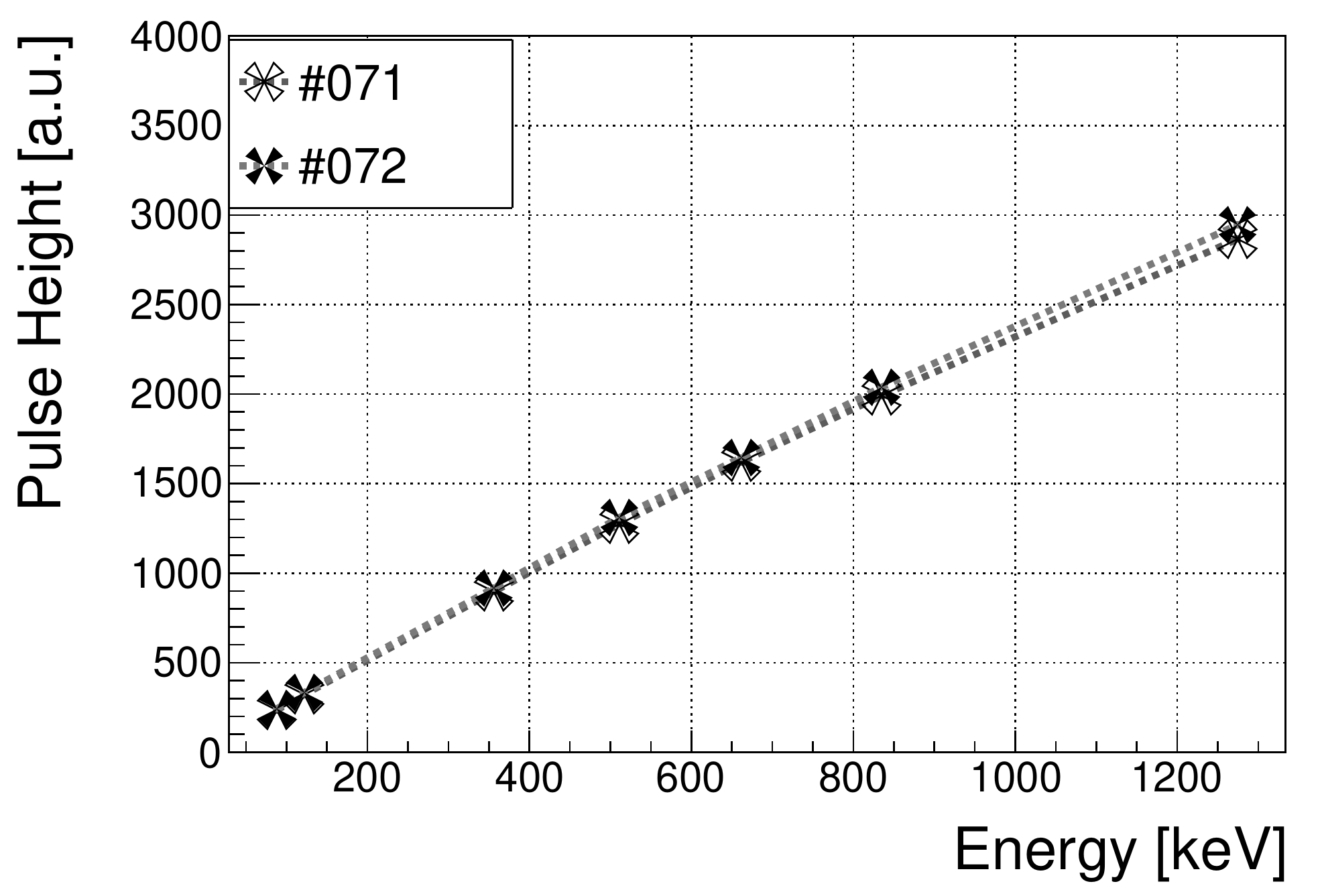}
\includegraphics[width=0.47\textwidth]{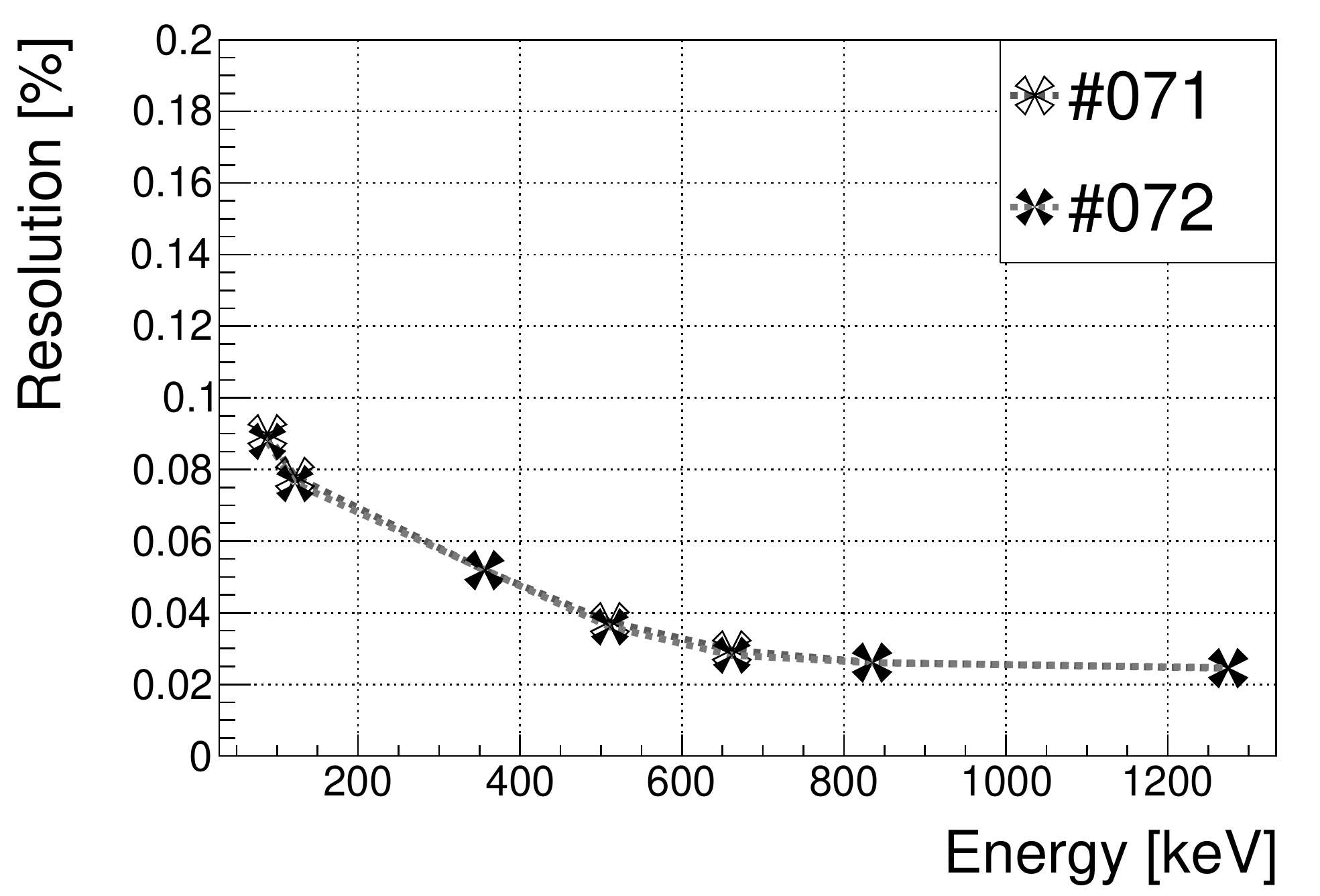}
\caption{Linearity (left panel) and energy resolution (right panel) dependence 
	for 1'' LaBr$_3$:Ce crystal read by Hamamatsu S14161-6050-AS SiPM
	arrays for two typical crystals.} 
\label{fig:crys2}
\end{figure}

Laboratory tests were done putting the detectors under test 
inside a Memmert IPV-30 climatic chamber, where
the temperature could be stabilized with a precision of $\sim$ 0.1 $^{\circ}$C.
Tests were done mainly at 25 $^{\circ}$C  in the following. Detectors were 
powered at their
nominal operating voltage $V_{op}$, as reported in table \ref{tab:sipm}. 
Different exempt X-rays sources (Cd$^{109}$, Co $^{57}$, Ba$^{133}$, Na$^{22}$, 
Cs$^{137}$, Mn$^{54}$) were used.
\begin{figure}[htbp] % figures (and tables) should go top or bottom of
                    % the page where they are first cited or in
                    % subsequent pages
\centering
\vskip -0.2cm
\includegraphics[width=0.44\textwidth]{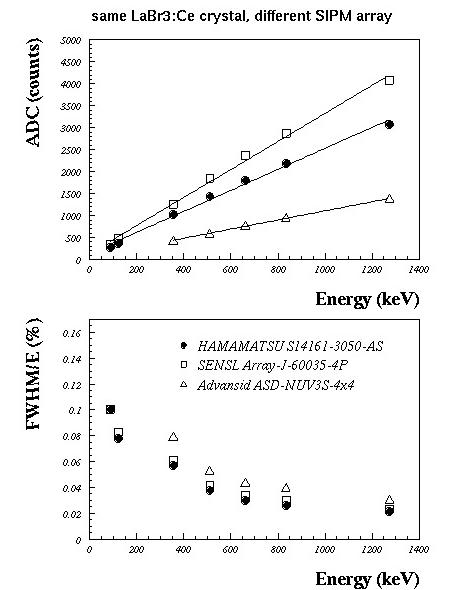}
\includegraphics[width=0.42\textwidth]{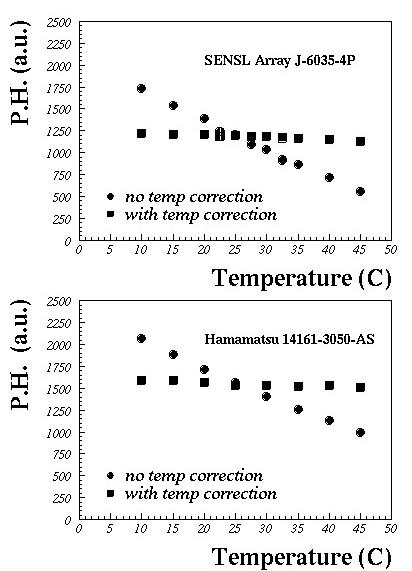}
\caption{Left panel: linearity (top) and energy resolution (bottom) dependence 
	for a 1/2'' LaBr$_3$:Ce crystal read by different SiPM arrays. 
	The same crystal was used in all tests. Right panel: P.H. response in
	a.u. for a typical 1/2" detector, with and without temperature 
	correction, with different SiPM array readouts.} 
\label{fig:crys}
\end{figure}
The summed analogue signal from the cells of a SiPM array is directly fed into a CAEN V1730 
fast digitizer (500 MHz bandwidth, 14 bit resolution) and is acquired by a custom
developed DAQ system \cite{ms}. Produced n-tuples are analyzed by the 
ROOT package 
\cite{root}. 
Figure \ref{fig:crys} reports linearity and FWHM energy resolution for readout
made with different SiPM arrays of the same LaBr$_3$:Ce crystal. 
Resolutions up to $3 \%$ 
at the Cs$^{137}$ peak are obtained with Hamamatsu or Sensl arrays. They compare well with what obtained with 
the best PMT readout. 
Detectors with SiPM array readout have typically longer risetimes ($\sim 20$ ns
for 20-80 $\%$ risetime)
as compared to ones with fast PMT readout. To reduce this problem, due to the
used parallel ganging of SiPM cells, an hybrid ganging solution is under study.
Figure \ref{fig:crys1} shows the results obtained with a sample of 
LaBr$_3$:Ce 1/2" 
crystals read by Hamamatsu S13161-3050 arrays.
Figure \ref{fig:crys2} reports instead linearity and FWHM energy resolution for two 
typical one inch LaBr$_3$:Ce crystals read by Hamamatsu S14161-6050-AS arrays.
In both cases resolutions up to $\sim 3 \%$ are obtained at the Cs$^{137}$ peak.
At the energies of interest for the FAMU experiment ($\sim 130$ keV) resolution
up to $\sim 7 \%$ are obtained. This compares well, with what obtained with a
conventional PMT readout, as shown in reference \cite{ad1}. 
\section{Performances in beam}
\begin{figure}[htbp] % figures (and tables) should go top or bottom of
                    % the page where they are first cited or in
                    % subsequent pages
\centering
\vskip -0.3 cm
	\includegraphics[width=\textwidth]{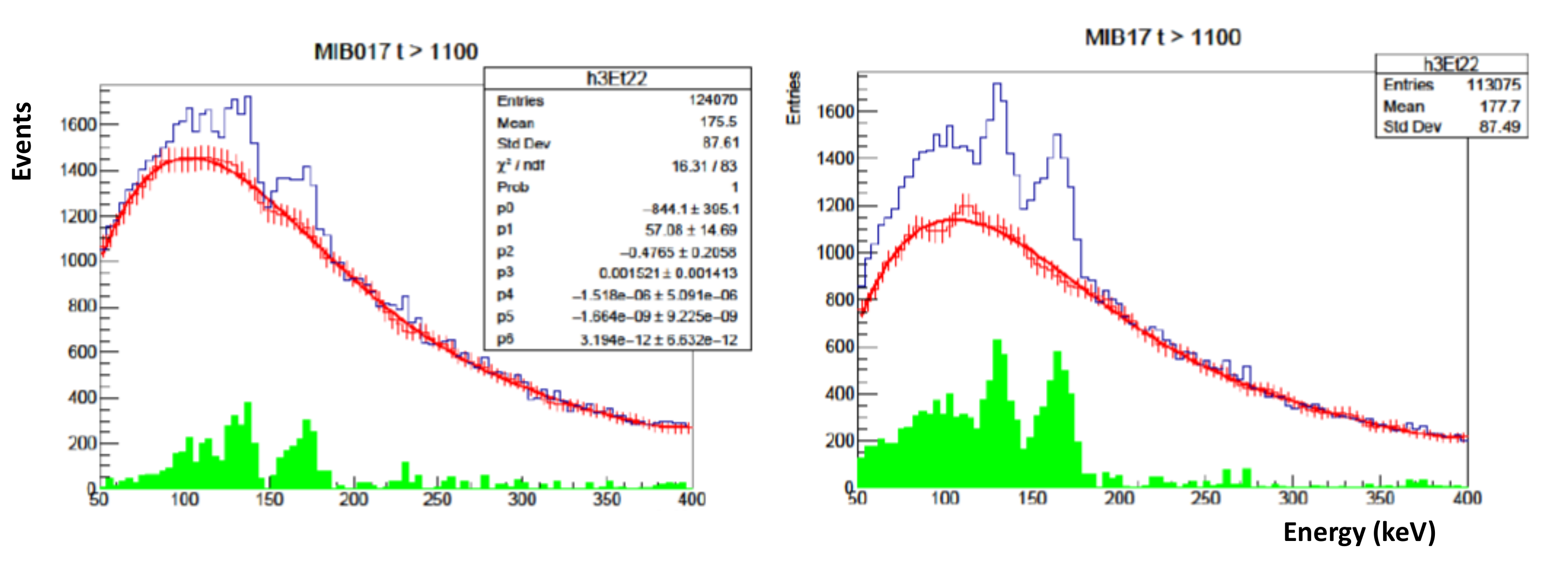}
\vskip -0.5 cm
\caption{Delayed spectra from a 1/2" LaBr3:Ce detector on the muon beam 
at RAL, using a target filled with a $H_2$+$O_2$ (0.3 $\%)$ mixture 
	at 7 bar (left panel) and 11 bar (tight panel), at 80 $^{\circ}$K. 
%%$K_{\alpha}$ and $K_{\beta)/K_{\gamma}$ lines are clearly visible, 
%%after background subtraction.
}
\label{fig:ral}
\end{figure}

In December 2018, eight 1/2" detectors were used for the FAMU data taking, 
where the RIKEN-RAL pulsed muon beam impinged on a target filled with a 
$H_2 + O_2 (0.3 \%)$ mixture at 7 or 11 bar pressure at 80 $^{\circ}$K.
After background subtraction (in red), signals corresponding to the Oxygen
$K_{\alpha}$ or $K_{\beta}$/$K_{\gamma}$ lines at $\sim$ 130 and 160 keV 
are clearly visible (in green) in the delayed spectra, reported in 
figure \ref{fig:ral} \cite{ghittori}.
The use of two crowns of LaBr$_3$:Ce crystals read by SiPM arrays is foreseen 
for the final FAMU data taking in 2022-2023, in addition to one equipped with
conventional PMTs.

\section{Conclusions}
Good performances and FWHM energy resolution up to $3 \%$ have been obtained
for both 1/2"  and 1"  LaBr$_3$:Ce crystals read by SiPM arrays. The gain
drift with temperature was corrected online by a custom developed NIM 
module, based on CAEN A7585D units.

\end{document}